\documentstyle[12pt]{article}
\begin{document}
\title{Description of Heavy Quark $\overline{MS}$ Mass by Lippmann Schwinger Equation}
\author 
{N. Tazimi \thanks{ nt$_{-}$physics @ yahoo.com}\\ M. Monemzadeh\thanks{monem@kashanu.ac.ir}\\
M. R. Hadizadeh\thanks{hadizade@ift.unesp.br}\\
\it\small $^{1,2}${{Department of Physics, University of Kashan, Iran.}}\\
\it\small$^{3}${{Instituto de Fısica Teorica, UNESP, 01405-900 Sao Paulo, Brazil}}}
\date{}
\maketitle

\begin{abstract}
Quark masses are of great prominence in high-energy physics. In this paper, we have studied the heavy meson systems via solving the Lippmann-Schwinger equation by using the Martin potential for heavy quark masses.
We have also attempted to use Martin potential to find an acceptable mass spectrum for heavy quarkonia. 
We obtained this spectrum via minimal phenomenological model (M. Melles, Phys. Rev. D \textbf{62}: 074019 2000).
The mass spectra for $b\bar{b}$ and $c\bar{c}$ are calculated without taking into account the relativistic corrections and spin-dependent effects. The obtained mass spectra turn out to fit the experimental findings. By using the conventional spectrum, we extract the pole mass of heavy quarks and use it along with the relation between $\overline{MS}$ mass (modified minimal subtraction scheme) and the on-shell quark mass to find $\overline{MS}$ mass for these quarks. The obtained results for $\overline{MS}$ mass are in good agreement with corresponding values reported in the literature.

\end{abstract}

\newpage
\section{Introduction}
QCD has two important characteristics. First, due to asymptotic freedom, quark interaction in high energy is weak. Second, since quarks do not exit in isolation, the inter-quark force in low energies increases with quark distances \cite{2}.
Because of the high masses of such quarks as $b$, $c$, $t$ in comparison to positronium, phenomenological potential models could be used for meson spectroscopy. Heavy quarkonium spectrum is a good means of investigating static interaction \cite{3} and the mass of heavy quark is an essential requirement for exploiting this investigation. 
The quark masses are of prevalent interest among high energy physicists, however, it is impossible to measure directly the heavy quark masses because of the confinement of the quarks within the hadrons. 
The heavy quark masses could be defined indirectly through analyzing their effects on the hadrons \cite{6}.\\
 

The precious information about the quark-antiquark ($q$-$\bar{q}$) interaction for ground and excited states of the mesons can be obtained by comparing the experimental data with the theoretical predictions \cite{2}.
In QCD Lagrangian, the quark mass parameters are not related to the quark masses which are used in the potential models. In these Lagrangians, therefore, a more common definition of mass which depends on the renormalization scheme is used.
Two different definitions of mass are often used in the renormalization of the QCD Lagrangian; $\overline{MS}$ mass and pole mass. The pole mass and the $\overline{MS}$ mass are renormalized quark masses in the on-shell renormalization scheme and in the modiﬁed minimal subtraction scheme, respectively. The mass definition that we use depends on the physical situation \cite{7}.
$\overline{MS}$ scheme is the most popular renormalization scheme for QCD perturbation. Non-perturbative effects at high energies or short distances are of little significance.
The most QCD calculations are performed by using $\overline{MS}$ scheme which is applicable for both light and heavy quarks. This scheme is also of use in estimating the quark masses in Standard Model. For all the above, $\overline{MS}$ scheme is considered as a perfect scheme for studying the physical properties of quarks. \\

The paper is organized as follow. In Sect. \ref{MS and Pole Mass} we explain $\overline{MS}$ and pole mass and the relation between these two masses. In Sect. \ref{Lippmann-Schwinger Equation} we present the non-relativistic Lippmann-Schwinger integral equation for two quark bound state by using a local interaction. In Sect. \ref{Potential and Mass Spectrum} we have introduced a potential model for heavy quark systems and we have presented our numerical results for the mass spectrum and $\overline{MS}$ mass for these systems. Finally a conclusion is given in the Sect. \ref{Conclusion}.

\section{$\overline{MS}$ and Pole Mass} \label{MS and Pole Mass}

In this paper we study the heavy quarks. For a heavy quark, $m\gg\Lambda_{QCD}$ \cite{8}, where $\Lambda_{QCD}$ is the scale of strong interaction in QCD, and its magnitude is approximately 200 MeV. We list the bare heavy quark masses used in this work in the Table \ref{table1}.

\begin{table}[h] 
\centering
\caption {The bare heavy quark masses. }
\begin{tabular} {ccccccccccc}
 \hline quark &&&&&&&&&  mass (GeV) \\\hline\hline
 {$ c $} &&&&&&&&&{$ 1.51$}\\ 
{$ b $} &&&&&&&&&  {$ 4.88 $} \\ 
{$ t $} &&&&&&&&& {$ 174 $} \\ \hline 
\end{tabular}
 \label{table1}
\end{table}

One of the results of analyzing the heavy quark spectrum is that the position of the pole in the propagator is considered as the pole mass \cite{7}. The difference between the quark mass and the pole mass is considered as an imaginary part (which is a multiple of $\Lambda_{QCD}$) \cite{9}:

\begin{eqnarray}
m_{Q} &\equiv& m_{pole}-\delta \hat{m}, \label{eq1} \\
\delta \hat{m}&=& i \, \delta m=i \, Im( m_{pole}), \label{eq2} \\
Im( m_{pole}) &=& const. \times \Lambda_{QCD}. \label{eq3}
\end{eqnarray}

 The relation between these two masses is:
\begin{equation}
m_{pole}=m_{\bar{MS}} [1+M(\bar{\alpha}_{s})],
\label{eq4}
\end{equation}
 where:
\begin{equation}
M(\bar{\alpha}_{s})=\sum_{n=0}^{\infty}P_{n} \bar{\alpha}_{s}^{n+1},
\label{eq5}
\end{equation}
where $P_{n}$ is a function of $n_{l}$ ($n_{l}$ is the quark flavor) and $\alpha_{s}$ is the strong interaction constant. For three-loop order, the relation is\cite{3,4,4-2,5}:

\begin{equation}
m_{Q}=\bar{m}_Q(\bar{m}_Q) \left [   1+ \frac{4}{3} \frac{\alpha_s  (\bar{m}_Q ) }{\pi}  +\xi_{2} \biggl( \frac{\alpha_s (\bar{m}_Q ) }{\pi} \biggr)^2 +\xi_{3} \biggl( \frac{\alpha_s (\bar{m}_Q ) }{\pi} \biggr)^3 \right],
\label{eq6}
\end{equation}

where $m_{Q}$ is the pole mass, and ${\bar{m}_{Q} (\bar{m}_{Q})}$ is the running mass in {$\overline{MS}$} scheme and $\xi_{2}$ and $\xi_{3}$ are a function of $n_{l}$ which are given explicitly in Ref. \cite{18} as:
\begin{eqnarray}
\xi_{2} &=& 13.44 - 1.041 \, n_{l}, \\
\xi_{3} &=& 194(5) -27.0(7) \, n_{l}+0.65 \, n_{l}^2.
\label{eq20}
\end{eqnarray}

If $m_{pole}$ is substituted in terms of $\bar{MS}$ mass, then the static energy of the quark-antiquark system could be \cite{10}:
\begin{equation}
E_{tot}(r)=2\,m_{pole}+V_{QCD}(r).
\label{eq7}
\end{equation}

 Of course, the inter-quark force could be used instead of the total energy because the perturbative extension of F(r) is far more convergent than the potential \cite{11}:
\begin{equation}
F(r)=-\frac{d}{dr}E_{tot}(r)=-\frac{d}{dr}V_{QCD}(r).
\label{eq8}
\end{equation}

\section{Lippmann-Schwinger Equation for Two-Body Bound State} \label{Lippmann-Schwinger Equation}

The bound state of two particles which interact by potential $V$ is described by homogeneous Lippmann-Schwinger equation:
\begin{equation} 
\left | \psi \right > =G_{0} \, V \left | \psi \right > ,       
\label{eq9}
\end{equation}
$G_0=\frac{1}{E_b-H_0}$ is the free two-body propagator, where $E_{b}$ is the binding energy of two-quark bound state and $H_0$ is free Hamiltonian.
In configuration space the equation (\ref{eq9}) turns out as:
\begin{equation} 
 \psi ({\bf r})=-m\sqrt{\frac{\pi}{2}}{\int d^3 r' } \frac{e^ {-\sqrt{m\, \left | E_{b} \right |} \, \left |{{\bf r}- {\bf r'} }\right | }}{\left |{{\bf r}- {\bf r'} }\right |} V(r' ){ \psi ({\bf r'} )},   
\label{eq10}
\end{equation}
where $m$ is the average mass of bare quarks. The compact form of the equation (\ref{eq10}) is:
\begin{equation}
{ \psi (r)}=\int_{0}^ {\infty}dr'  M(r,r') \, { \psi (r' )},    
\label{eq11}
\end{equation}
 where:
\begin{equation} 
M(r,r')=-\sqrt{2} \, m \, \pi^{3/2} \, \int_{-1}^ {1} dx' \, \frac{e^{-\sqrt{m \, \left | E_b \right | } \, \sqrt{r^2+r'^2-2r r' x' }   }}{\sqrt{r^2+r'^2-2rr' x' } } \\ 
r'^2 \, V(r').
\label{eq12}
\end{equation}

The eigenvalue integral equation (\ref{eq11}) can be written schematically as:
\begin{equation}
K(E) \left | \psi \right > =\lambda(E) \left | \psi \right > ,
\label{eq13}
\end{equation}
where $K(E)$ stands for the kernel of eigenvalue equation, which is energy-dependent. $\lambda(E)$ is the eigenvalue, and $\left | \psi \right > $ denotes the two-body wave function. For a physical solution of the eigenvalue equation \ref{eq13} and to obtain the binding energy of the system, i.e. $E=E_b$, one should solve this equation to obtain the eigenvalue $\lambda=1$. The integral equation \ref{eq11} can be solved by iteration or direct methods. Since there is not any shifted argument in the wave function amplitude, i.e. $\psi (r' )$, it would be easier to solve this integral equation directly, without using any iteration procedure. In order to solve the integral equation \ref{eq11}, we should first discretize the continuous configuration and angle variables \cite{12}. To this aim, for discretization of both configuration and angle variables we use the linear mappings by using the Gaussian quadrature grid points, where $r_i$, $r'_j$ and $x'_k$ are the mesh points corresponding to $r$, $r'$ and $x'$ variables with number of mesh points $N_r$, $N_{r'}$ and $N_{x'}$, correspondingly. By these considerations, the eigenvalue equation \ref{eq11} can be written as: 
\begin{equation}
{ \psi (r_i)}=\sum_{j=1}^{N_{r' }} \, W_{r'_j} \, M(r_i,r' _{j}) { \psi (r' _{j})},
\label{eq14}
\end{equation}
where the matrix elements of the kernel $M(r_i,r' _{j})$ can be obtained from equation \ref{eq12} as:
\begin{equation} 
M(r_i,r'_j)=-\sqrt{2} \, m \, \pi^{3/2} \, \sum_{k=1}^{N_{x'}} \, W_{x'_k} \, \frac{e^{-\sqrt{m \, \left | E_b \right | } \, \sqrt{r_i^2+r_j'^2-2r_i r'_j x'_k }  }}{\sqrt{r_i^2+r_j'^2-2r_i r'_j x'_k } } \,
r_j'^2 \, V(r'_j),
\label{eq15}
\end{equation}
where $W_{r'_j} $ and $W_{x'_k}$ are the point weights of Gauss-Legendre polynomials. 
In the last step by diagonalization of the kernel of eigenvalue equation:
\begin{equation} 
K(r_i,r'_j)=W_{r'_j} \, M(r_i,r'_j),
\label{eq16}
\end{equation}
one can obtain the binding energy of two-quark system for each energy level.


\section{Potential and Mass Spectrum} \label{Potential and Mass Spectrum}

 Many potential models have been proposed for inter-quark interaction. Some of them suit for light hadrons, and some other are suitable for heavy hadrons. For example, the models proposed by A. De R\'ujula et al. are not used to describe the light-quark systems \cite{13}, Martin models apply well to nonrelativistic calculations of heavy mesons \cite{14} and Cornel potential is appropriate for heavy and light hadron systems \cite{15,15-2}.

 In this paper, we have used the Martin potential (low-power potential) to study the mass spectrum of heavy mesons such as $b\bar{b}$ and $c\bar{c}$ systems \cite{14}. 
The functional form of the Martin potential is given by:

\begin{equation}
V(r)=b_{m}+a_{m}(c_{m}\,r)^{0.1},
\label{eq17}
\end{equation}
where the parameters of potential are:
\begin{equation}
b_{m}=-8.064 \, GeV,\ \ \ a_{m}=6.898\, GeV,\ \ \ c_{m}=1\, GeV.
\label{18}
\end{equation}

Accordingly, in the first step, i.e. estimating the binding energy for bottomonium and charmonium, has turned out to submit reasonable findings based on the mass formula.
We have shown our results for the mass spectrum of bottomonium and charmonium systems in Tables 2 and 3 in comparison to other theoretical results and experimental data.

\begin {table} [h] 
\centering
\caption{Calculated bottomonium (b\={b}) mass spectrum (in unit GeV) for Martin potential and 
for spin=1.}
\begin{tabular}{ccccccccccccc}
\hline state &&&& Ref. \cite{16} &&&& our work &&&& experiment \cite{17} \\ \hline \hline
1S &&&& {$ 9.46 $}&&&&{$ 9.45 $} &&&&{$ 9.460 $} \\
2S	 &&&&{$ 10.03 $}&&&& {$ 10.051 $}&&&&{$10.023 $} \\
3S &&&& {$ 10.36 $}&&&&{$ 10.32 $}	&&&&{$ 10.355 $} \\ 
1P &&&&{$9.90 $} &&&&{$ 9.89 $} &&&& {$ 9.9$}\\ 
2P &&&& {$ 10.26 $}&&&&	{$ 10.249 $} &&&&{$ 10.26 $} \\
1D &&&&{$ 10.15 $}&&&& {$ 10.169 $}	&&&&{$ 10.161 $} \\ \hline
\end{tabular} 
\label{table2}
\end{table}

\begin {table}[h] 
\centering
\caption {Calculated charmonium (c\={c}) mass spectrum (in unit GeV) for Martin potential and 
for spin=1.}
\begin{tabular}{ccccccccccccc }
\hline  state  &&&& Ref. \cite{16} &&&& our work &&&& experiment \cite{17} \\\hline
1S  &&&& {$ 3.097 $} &&&& {$ 3.07 $}  &&&& {$ 3.068 $} \\
2S	 &&&& {$ 3.69 $}&&&& {$ 3.75 $} &&&&{$ 3.672 $} \\
3S &&&& {$ 4.78 $} &&&& {$ 4.09 $}	 &&&& {$ 4.040$} \\ 
1P  &&&& {$ 3.528 $} &&&& {$ 3.528 $} &&&& {$ 3.525$}\\ 
2P &&&& {$ 3.944 $}&&&&	{$ 3.963 $} &&&& {$ - $} \\
1D  &&&&{$ 3.806$} &&&& {$ 3.803 $}	&&&& {$ 3.779 $} \\ \hline
\end{tabular} 
\label{table3}
\end{table}

In order to be able to calculate the running mass in $\overline{MS}$ scheme for heavy quarks from equation \ref{eq6}, we need to extract the pole mass $m_{pole}$ from the calculated mass spectra, our results for the pole masses are as:

\begin{equation}
\begin{array}{l}
m_{b}(2-loop)\simeq 4.72 \, GeV, \\
m_{c}(2-loop)\simeq 1.4 \, GeV.\\
\end{array}
\label{eq19}
\end{equation}

By using the following values for coefficients $\alpha_{s}$ and $\xi_{2}$, which are taken from Ref. \cite{3}, for $b$ for $c$ quarks:

\begin{eqnarray}
\alpha_{s}(\bar m_{b}) &\simeq& 0.217 \nonumber \\
\xi_{2} &\simeq& -9.27
\label{eq21}
\end{eqnarray}

\begin{eqnarray}
\alpha_{s}(\bar m_{c}) &\simeq& 0.31 \nonumber \\
\xi_{2} &\simeq& -10.31
\label{eq22}
\end{eqnarray}
and by considering the mass poles given in equation \ref{eq19}, we can obtain the $\overline{MS}$ masses, which are given in Table \ref{table4}.
The comparison of our numerical results with the values reported by the Particle Data Group (PDG) \cite{19,21}, indicates that the two sets of results are in good agreement .
\begin {table}[h] 
\centering
\caption {The running mass in $\overline{MS}$ scheme for heavy quarks.}
\begin{tabular}{ccccccccc }
\hline
 meson  &&&& Refs. \cite{19,21} &&&& our work \\\hline \hline
$\overline{m_{b}}(\overline{m_{b}})(2-loop)$ &&&&{$ 4.19 $} &&&&{$ \simeq 4.33 $} \\
$\overline{m_{c}}(\overline{m_{c}})(2-loop)$ &&&&{$ 1.23 $}&&&&{$ \simeq 1.35 $} \\
$\overline{m_{t}}(\overline{m_{t}})(2-loop)$ &&&&{$ 166 $}&&&&{$ \simeq 166.36 $} \\\hline
\end{tabular} 
\label{table4}
\end{table}


 The top quark mass is an important parameter in standard model at high energies \cite{20}. Top quark has a width of $\Gamma\cong1.5\, cm $, which is larger than $\Lambda_{OCD}$. Because of this short lifetime, top quark decays very quickly. $\overline{MS}$ mass, which is a short-distance mass, could be determined very accurately \cite {21}.
For $t$ quark, we follow the same procedure. However, a mass spectrum does not exist for this unstable meson by having a short lifetime \cite{21}. 

 

 We have solved the Lippmann-Schwinger equation for this meson by using the Martin potential and we have shown our numerical results for $t\bar{t}$ binding energy for both spin 0 and 1 states in table (\ref{table6}).\\

\begin {table} [h]
\centering
\caption{t\={t} binding energy for both s=0 and 1 states. The parameter A denotes the spin coupling coefficient. The stability interval \cite{12} is related to the configuration cutoff which has been considered in numerical solution of the eigenvalue integral equation \ref{eq10}. }
\begin{tabular}{ccccccccccccc }
\hline spin &&& {$A[MeV^{3}/c^{6}]\times10^{10}/h^2$}  &&& {$ BE \, [MeV]$}  &&& stability interval $[fm^{-1}]$ \\\hline \hline
s=1  &&& {$-449.01 $}  &&& {$-11.296 $}  &&& {$ [0.3,0.4],[0.44,1.2] $} \\
s=0  &&& {$45.759 $}  &&& {$ -36.948 $}  &&& {$ [0.3,0.4],[0.43,1.1] $} \\ \hline
\end{tabular} 
 \label{table6}
\end{table}

By using the following values for the coefficients $\alpha_{s}$ and $\xi_{2}$ \cite{22}:
\begin{eqnarray}
\alpha_{s}(\bar{m_{t}}) &=& 0.1085 \nonumber \\
\xi_{2} &\simeq& -10.31
\label{eq23}
\end{eqnarray}
we obtain the $\overline{MS}$ mass for $t$ quark which is given in Table \ref{table4} and is in good agreement with corresponding values reported  in Refs. \cite{19,21}.

We would like to add the comment that the integral equation \ref{eq10} is singular for the confining potentials, and consequently the calculated energy eigenvalues would not be in agreement with the exact analytic binding energies.
To overcome this problem one can use a regularized form of the confining potentials to remove
the singularity of the kernel. To this aim one can keep the divergent part of the potential
constant after exceeding a certain distance, which creates an artificial barrier. So, we have fixed the potential in the stability interval and our numerical calculations show that the physical eigenvalue $\lambda=1$ is quite stable in this region for configuration cutoffs.

 \section{Conclusion} \label{Conclusion}
 We have used the Martin potential to study the heavy quark systems by solving the homogeneous Lippmann-Schwinger equation without considering the relativistic corrections and also the spin-dependent effects. We have obtained the mass spectrum for the heavy mesons and our numerical results are in good agreement with other theoretical results and experimental data as well. 
We have calculated the pole mass by fitting the mass spectrum of $b$ and $c$ quarks to other references. We have specified the running mass $\overline{MS}$ for these systems. Our obtained masses are close to the findings reported by the Particle Data Group.
For $t$ quark, we have verified the stability interval of regularized form of the Martin potential.

\section*{Acknowledgments}
M. R. Hadizadeh acknowledge partial financial support from the Brazilian agency
Funda\c c\~ao de Amparo \`a Pesquisa do Estado de S\~ao Paulo (FAPESP).


\end{document}